\documentclass[a4paper,fleqn,usenatbib,referee,epsfig,color]{mn2e}
\usepackage[T1]{fontenc}
\usepackage{ae,aecompl}

\usepackage{graphicx}	% Including figure files
\usepackage{amsmath}	% Advanced maths commands
\usepackage{amssymb}	% Extra maths symbols

\newif\ifAMStwofonts
\newcommand{\be}{\begin{equation}}
\newcommand{\ee}{\end{equation}}
\newcommand{\bea}{\begin{eqnarray}}
\newcommand{\eea}{\end{eqnarray}}

\begin{document}
\title[Trans-Light Transformations in Special Relativity ]{\bf How Might Tachyons Appear? }
\author[R.N. Henriksen$^1$ ]
{R. N. Henriksen$^1$\thanks{henriksn@astro.queensu.ca} \\ 
%R. Beck$^2$\thanks{rbeck@mpifr-bonn.mpg.de}, R. Walterbos$^3$\thanks{rwalterb@nmsu.edu}, and
%Q. D. Wang$^4$\thanks{wqd@astro.umass.edu}\\
$^1$Dept. of Physics, Engineering Physics \& Astronomy, Queen's University, Kingston, Ontario, K7L 3N6, Canada\\}
\date{Accepted XXX. Received YYY; in original form ZZZ}
\pubyear{2020}

%\address{$^1$
%R. Beck$^2$\thanks{rbeck@mpifr-bonn.mpg.de}, R. Walterbos$^3$\thanks{rwalterb@nmsu.edu}, and
%Q. D. Wang$^4$\thanks{wqd@astro.umass.edu}\\
%Dept. of Physics, Engineering Physics \& Astronomy, Queen's University, Kingston, Ontario, K7L 3N6, Canada}
%\date{Accepted XXX. Received YYY; in original form ZZZ}
%\pubyear{2017}
% Don't change these lines
%\begin{document}
\label{firstpage}
\pagerange{\pageref{firstpage}--\pageref{lastpage}}
\maketitle
%%% Subject entries to be placed here %%%%
%\subject{Supra Luminal Lorentz}

%%%% Keyword entries to be placed here %%%%
%\keywords{Light Horizon,Tachyons, Observers }

%%%% Insert corresponding author and its email address}
%\corres{R.N. Henriksen\\
%\email{henriksn@astro.queensu.ca}}
\begin{abstract}
Assuming the existence of supra luminal matter,  referred to as `tachyonic', we reconsider possible Lorentz style transformations between tachyon observers and sub luminal (`bradyons')  observers. We consider a unique possibility following from a straightforward argument based on relative motion as 
a Lie group. The result is  novel in that it requires the time direction to be reversed for tachyon observers. We use this result to find the transformation between supra luminal observers.
%The covering is humorously described for brevity as `condo coordinates'. 
%Our principal innovation is to recognize an interchange of space-like and time-like variables in the `plane' of coordinate time and the direction of  relative motion, on crossing the horizon.  

%We give the transformations between supra luminal observers for completeness.%This is the beginning of the `warp' in our title. 
An extended discussion {\it speculates} concerning  physical  evidence for, and consequences of,  a supra luminal regime dual to the sub luminal regime.  It appears that supra luminal particles are likely to be of very low energy and hence be difficult to detect. However their momentum may be significant depending on their asymptotic mass.  Tachyons are candidates for astronomical `dark matter' and perhaps  vacuum energy as manifested in the cosmological constant.  Quantum tachyons might be detected as periodic variations in Casimir type measurements corresponding to their De Broglie wavelength.%This is the `duality' in our title. 
%The duality also appears when the De Broglie wave of a particle is considered, which is supra-luminal when the particle is sub-luminal and vice versa. 
  We suggest  that supra luminal and sub luminal particles can be entangled  at  both Cauchy and Event  horizons, so that transitions may be possible for quantum particles.

%Quantum uncertainty might allow for transitions at the horizon,
%Cauchy horizon.
\end{abstract}
\maketitle
%\begin{fmtext}
\keywords{Trans-luminal  Lorentz transformations, Tachyon measurement, Dark Matter }
%\end{keywords}
\newpage

\section{Introduction}
There is a recognizable longing expressed through our collective imagination for faster than light phenomena and ultimately for faster than light  travel capability. For this reason alone one must venture intellectually  with extreme caution and humility in this direction. I will not argue formally for the existence or implications of faster than light objects, except by a few manifest speculations in the discussion.  I do wish to offer a more direct route to the Lorentz transformations as extended to faster than light relative velocity. This approach is  based on  the methods described in \cite{CH1991} and \cite{Hen2015}. % Our results  help to choose between the possibilities  presented  by other authors \cite{BDS1962}, \cite{HC2012}, but with some differences of interpretation.  

These  collective imaginings have mostly been the province of science fiction, but scientific speculation has also played a r\^ole. This speculation  seems to have been revitalized in `modern' times in \cite{BDS1962}, but more recently one finds new discussion in \cite{HC2012}, \cite{HC2014} , \cite{V2012} plus their included references and related articles.  Earlier references together with an (optimistic) survey of possible supra light phenomena may be found in \cite{Rec2001}.

%In this article I will restrict speculation severely in favour of rigorous deduction, but there is a sense in which I can  illustrate the `warp drive'  \cite{Al1994} and wave/particle duality.  I hope these will be regarded in the light-hearted way intended, in the hope that someday, somewhere, they may strike a spark.

  I arrive at the principal result using a Lie symmetry of flat space-time.  This symmetry expresses the equality of  observers  moving with constant velocity in flat space-time, independent of the magnitude of this velocity. A discussion of the velocity transformation between supra luminal and sub luminal observers, and between supra luminal observers is included. Finally there is a section on discussion and conclusions in which some  considerable amount of speculation is allowed.

\section{Lorentz transformation as Lie motion}
\label{sect:supraLorentz}

In \cite{CH1991} and \cite{Hen2015}  `scale invariance' or `self-similarity' was treated as a Lie motion along a certain direction in parameter (including the measures of space and time) space. The invariants {\it under this motion} are the  `variables' on which a scale invariant system may depend. In particular in \cite{Hen2015} (pp 96-98) it was shown that in Minkowski space-time (reduced  Galilean parameters $\{x,t\}$), uniform relative motion can be regarded as such a Lie motion. Moreover it was found that the Lie invariants  were the co-moving Galilean coordinates of one observer $O'$, seen as moving with speed $u$ along the $x$ axis of observer $O$. Solving the Lie equations that determine the invariants  along the space-time direction ${\bf k}$ of the motion, yielded the Lorentz transformations  between the coordinates of O and the co-moving coordinates of $O'$. This approach is similar to, but a little more formal, than that used in \cite{HC2012}. It was performed independently.

 This solution for the invariants $\{x',t'\}$  (and therefore the coordinates of $O'$) was found to be (\cite{Hen2015},(3.54)) 
 \bea
 x'&=&x\cosh(\alpha uT)-t\sinh(\alpha uT),\label{eq:Liex'}\\
 t'&=& t\cosh(\alpha uT)-x\sinh(\alpha uT),\label{eq:Liet'}
 \eea
 where $T$ is the Lie parameter parallel to the motion such that $\partial_T\propto {\bf k}$ is the Lie derivative along the motion. It is a  logarithmic measure of the $O$ observer's time. The quantity $1/\alpha$ is a convenient Unit of time that is a known function of $u$. Constants have been chosen so that $x'=x$ and $t'=t$ at $t=0$. Thus the motion starts from an identity as required for the Lie group motion, and indeed is also required by the Lorentz transformations in standard form. These relations are valid for any $T$  as the relative motion along the $x$ axis continues. However at any fixed $T$ the transformations are functions only of $u$ and give the  relations between the invariant coordinates of $O'$ and those of O. We choose $T=1$ to simplify the algebra. 

In the familiar sub luminal relative case, O tracks the observer $O'$  by following a fixed spatial point in $O'$ so that $dx'=0$.  Then equation (\ref{eq:Liex'}) shows that $\tanh(\alpha u)=u$. Subsequently we substitute for $\cosh(\alpha u)$ and $\sinh(\alpha u)$ with the usual sign choices into equations (\ref{eq:Liex'}) and (\ref{eq:Liet'}) to obtain the forward Lorentz transformations as 
 \bea
 x'&=&\frac{x-ut}{\sqrt{1-u^2}},\label{eq:FLx}\\
 t'&=&\frac{t-ux}{\sqrt{1-u^2}}.\label{eq:FLt}
 \eea
 The backward transformations follow as usual by changing the sign of $u$ and interchanging the primes.  The quantity $\alpha(u)$ also follows necessarily from $\tanh(\alpha u)=u$. It turns out \cite{Hen2015} to be simply related to a D\"oppler factor $K=\sqrt{(1+u)/(1-u)}$ ($\equiv t/t'$ for events along a light ray), as $\exp(\alpha u)=K$.

  However, our question now is what is the form of the invariants (i.e. the coordinates of $O'$) if we allow $u$ to become supra luminal?  That is if we allow $O'$ to cross what is a Cauchy horizon in flat space-time for the origin event. The solution for the invariants in equations (\ref{eq:Liex'}) and (\ref{eq:Liet'}) show no evident infinity or change in spatial-temporal significance when $u\ge 1$ The Cauchy horizon is hidden in the behaviour of $\alpha(u)$ as we show below.

 We know that using the preceding procedure  for the sub luminal relative motion will produce imaginary quantities through $\sqrt{1-u^2}$ on crossing the Cauchy horizon. To avoid this, let us first  rewrite the Lie equations (\ref{eq:Liex'}) and (\ref{eq:Liet'}) in the following differential forms ($u=dx/dt$)
 \bea
 dx'&=&dt\cosh(\alpha u)(u -\tanh(\alpha u)),\label{eq:diffFLx}\\
 dt'&=&dt\cosh(\alpha u)(1-u\tanh(\alpha u)).\label{eq:diffFLt}
 \eea
 Now if $u>1$ we must choose $\alpha(u)$ such that  $\tanh{(\alpha u)}=1/u$, because  then $cotanh{(\alpha u)}$ has the necessary supra luminal  range. \footnote{Accepting this we find that again  $\exp{\alpha u}=(\sqrt{(u+1)/(u-1)}$  so that $\alpha u\rightarrow 0$ as $u\rightarrow \infty$.} This implies, according to the second of these last equations, that  $O$  can only follow the observer in supra-luminal motion  by fixing $dt'=0$. {\it That is  by fixing a point in $t'$ as though it were space like}.  Even though during this flat space Lie motion there is no gravitational field, we find  nevertheless   that the  Cauchy trajectory $u=1$ retains a property of an event horizon. That is, {\it the physical significance of the dimensions  $x'$ and  $t'$ are interchanged}. We shall finally interchange also the labels, but we continue with the argument for the moment. 
 
 In summary equation ((\ref{eq:Liet'}) with $dt'=0$) instead of $dx'=0$ is compatible with the supra luminal condition  
 \be
 \tanh(\alpha u)=1/u,\label{eq:uLiesup}
 \ee
 From this  condition we obtain 
 \be
 \cosh^2(\alpha u)=\frac{u^2}{u^2-1},~~~~~~\sinh^2(\alpha u)=\frac{1}{u^2-1},\label{hyperbolicsLie}
 \ee      
 where the two sign choices must be the same for positive $u$. Consequently we obtain from equations (\ref{eq:Liex'}) and (\ref{eq:Liet'})
 \bea
 t' &=&\frac{ut-x}{\sqrt{u^2-1}},\nonumber \\
 x' &=&\frac{ux-t}{\sqrt{u^2-1}}.\nonumber
 \eea

However $t'$ is apparently space-like  because it identifies the moving position of $O'$, the orthogonal  $x'$ is therefore time-like,. Consequently we interchange  the labels $x',t'$ to obtain our transformations to supra luminal coordinates as
\bea
x'&=&\frac{ut-x}{\sqrt{u^2-1}},\label{eq:SL1}\\
t'&=& \frac{ux-t}{\sqrt{u^2-1}}.\label{eq:SL2}
\eea
These should be compared to the `evident' possibility that requires only reversing the sign of $1-u^2$ in the normal Lorentz transformations\cite{BDS1962}.

 The inverse equations are found by direct calculation and become
 \bea
 x&=&\frac{x'+ut'}{\sqrt{u^2-1}},\label{eq:SLB1}\\
 t&=&\frac{t'+ux'}{\sqrt{u^2-1}}.\label{eq:SLB2}
 \eea
 The reason for not imposing the pure relative definition of the motion (i.e. reversing the sign of $u$ and exchanging the primes)  in order to find the inverse transformations is because of the Cauchy horizon between the two observers. This creates an asymmetry between the two observers that allows time to run in the normal sense for the sub luminal observer, at the cost of allowing It to run in the opposite sense  ($dt'=-dt/\sqrt{u^2-1}$) for the supra luminal observer. Hence the sub luminal observer's past is the supra luminal observers future and vice versa. Similarly spatial ordering is reversed for the supra luminal observer.
 
 We note that there is a kind of rotation or `warping' of flat space-time on crossing the Cauchy horizon according to these transformations. We can see this by writing our equations slightly differently.
  We note that  the function $\alpha(u)$ can be written in the form ($u>0$)
 \be
 \alpha ~u=\ln{\sqrt{\frac{1+u}{|u-1|}}},\label{eq:alpha}
 \ee
 which holds in both the sub luminal and super luminal domains. 
 The equivalent complexified statement of the moving coordinates (Lie invariants)  given in equations (\ref{eq:Liet'}), (\ref{eq:Liex'})  are rotations in the $\{it,x\}$ plane 
 \be
 t'=x\cos(i\theta)+it\sin(i\theta),~~~~~~x'=t\cos(i\theta)+ix\sin(i\theta),\label{eq:complexM}
 \ee
 and the inverse transformations are found by reversing the sign of $\theta\equiv \alpha u $ and interchanging the primes.  The magnitude sign in equation (\ref{eq:alpha}) allows the transformations to be used for both sub-luminal to supra-luminal motion., although the infinity on the horizon can not be avoided. It is easy to verify  form these rotations that  $x'^2-t'^2= t^2-x^2$, so that finally the  axes are completely rotated  to preserve the metric signature.  

It is not yet evident what  preferred value of $u$ is appropriate for a  tachyon observer $O'$, who is presumably attached to a tachyon particle.  For a sub luminal observer, $u=0$ gives the minimum energy and zero momentum. Assuming that a tachyon has a classical `action' implies that the tachyon has zero energy (but finite momentum) as $u\rightarrow \infty$. Zero energy may be a preferred state (unless there are negative rest mass tachyons). 
Either set of equations (\ref{eq:SL1}), (\ref{eq:SL2}) or (\ref{eq:SLB1}), (\ref{eq:SLB2}) show that in this limit the rotation is complete because $x\rightarrow t'$ and $t\rightarrow x'$.  Relation  (\ref{eq:SLB1} implies that the infinite velocity tachyon is everywhere for $O$ with no elapsed time according to  relation (\ref{eq:SLB2}). This limit of infinie $u$  may in some sense be considered the `rest' limit of the tachyon. 
 
 Turning to practical matters,  time dilation for the $O$ observer relative to the $O'$ observer remains $dt=\gamma_+ dt'$ but that  for the $O'$ observer $dt'=-\gamma_+ dt$ (where we set $\gamma_+=1/\sqrt{(u^2-1)}$). The Lorentz-Fitzgerald contraction however is $dx=-dx'/\gamma_+$ for $O$ but rather $dx"=dz/\gamma_+$ for $O'$. The minus sign is somewhat unexpected! 
 
 However given the time reversal for $O'$, these reversed effects may be understood  diagramatically as  being due to the supra luminal relative motion combined with only luminal signal speed. We must carefully (allowing for $u>1$) plot a  supra luminal rod in the first quadrant  of the axes of $O$. This requires representing the axes of $O'$ in the first quadrant of the $O$ world and hence representing the moving rod as extended in time for $O$ because it lies along the $t'$ axis. Subsequently we propagate light waves from the ends of the rod  forward in $O$ time to intersect the $x$ axis of $O$. The consequent spatial order for $O$ relative to a common origin, is seen to be reversed relative to that fixed in the frame of $O'$. 

Similarly we can plot the axes of $O$ and a moving $O$ rod in the world of $O'$. These appear in the third quadrant to indicate correctly  the time reversal between $t$ and $t'$  and the negative relative motion. Drawing light rays from the ends of the rod  lying along the $t$ axis forward in $t'$ and backward in $t$  until an intercept with a $x'$ axis  is reached in each case, shows the order in space NOT to be reversed.

The D\"oppler effect  of a radiating tachyon source  (that is, not Cerenkov emission) becomes
\be
\nu_o=\frac{\nu_s}{\gamma_+|u\cos{\theta}-1|},\label{eq:doppler}
\ee
where $\nu_o$ and $\nu_s$ are observed and source frequencies from a source moving at an angle $\theta$ to the line of sight  directed towards the observer. So long as both $u$ and $|u\cos{\theta}|$ are large compared to $1$ this becomes $\nu_o=\nu_s/|\cos{\theta}|$, which yields a blue shift except for motion parallel to the line of sight. Should $|u\cos{\theta}|<1$ even though $u>>1$ then the blue shift is dominated by the transverse effect and can be very large for perpendicular motion ($|\cos{\theta}|<1/u$). 

This behaviour has an effect on the apparent transverse motion of jetted astronomical sources on the sky. This velocity is given by (e.g. \cite{Rees66},\cite{Hen2011}) 
\be
u_{app}=\frac{u\sin{\theta}}{|u\cos{\theta}-1|},
 \ee
 which gives $u\tan{\theta}$ for an approaching source with $u\cos{\theta}>>1$. This is only supra luminal for $\theta>\pi/4$.  A `blazer' type supra luminal source advancing along our line of sight gives the frequency shift $\nu_o/\nu_s=\sqrt{(u+1)/(u-1)}$. This goes to one in the very supra luminal limit.
 
 Another interesting property of a tachyon source that actually passes an observer, is that it continues to be `seen' (i.e. by radiated light) with a blue shift as it recedes in both directions for some limited time. This is because light rays emitted before the tachyon passes continue to arrive after the passage of the tachyon, and so also  do the light rays emitted  by the tachyon after the passage. Ultimately this phenomenon is limited in time by the flux density sensitivity of the observer and the speed of the tachyon.  This assumes radiation to be emitted in all directions. Cherenkov radiation from the tachyon  emitted in a forward cone would not be seen after the passage of the tachyon.
 
 It is easily checked that the parallel  velocity transformation is maintained by the transformations (\ref{eq:SL1}) and (\ref{eq:SL2}) in the standard form 
 \be
 \frac{dx'}{dt'}=\frac{dx/dt-u}{1-u(dx/dt)},\label{eq:vtrans}
 \ee
 and equations (\ref{eq:SLB1}) and (\ref{eq:SLB2}) yield the standard inverse result
 \be
 \frac{dx}{dt}=\frac{dx'/dt'+u}{1+u(dx'/dt')}.\label{eq:Inveltrans}
 \ee
 
 Maintaining this relation has been a requirement used by other authors to obtain trans luminal transformations(\cite{HC2012}). The behaviour is rather different for $u>1$ however. In particular if a tachyon observer measures a velocity  $dx'/dt'$ such that $u(dx'/dt')>>1$ then we measure a velocity $1/u+1/(dx'/dt')$, which for very large $u$ is just the reciprocal of the tachyon velocity for the supra luminal observer.

 \section{Transformations between two supra luminal observers}
 \label{sect:suprasupratrans}
 
 Although we can not share their perspective, it is of interest when considering interaction between tachyons to consider how they view one another. To this end we consider briefly how to extend the supra light transformations to relatively moving tachyon observers.
 
 We consider two supra-luminal observers $O_1$ and $O_2$ moving respectively with speeds $u_1$ and $u_2$ along the x axis of a sub luminal inertial observer $O$.  We may think of the ordering along $x$ as $\{O,O_1,O_2\}$ with $u_2>u_1$ for definiteness, but we are not restricted to this ordering. The equations (\ref{eq:SLB2}) and (\ref{eq:SLB1}) may be written as the transformations between $t$ and $x$ and each of the sets of  supra luminal coordinates $\{t_1, x_1\}$ and $\{t_2 , x_2\}$ respectively.  Equating the two expressions for $t$ and the two expressions for $x$  allows one set of supra luminal coordinates to be solved for in terms of the other. This gives the transformations between supra luminal observers  as
 \bea
 t_1&=&\gamma_1\gamma_2\left((u_1u_2-1)t_2-(u_2-u_1)x_2\right),\nonumber\\
 x_1&=&\gamma_1\gamma_2\left((u_1u_2-1)x_2-(u_2-u_1)t_2\right),\label{eq:betu1u2}
 \eea
 where  the form of $\gamma\equiv 1/\sqrt{u^2-1}$ and the sub-scripts correspond to the subscripts on $u$.  The inverse transformations are found by interchanging the sub-scripts $1$ and $2$. 
 
  The supra luminal observer $O_1$ may obtain the proper time of $O_2$  in relation to the coordinate time $dt_1$ by setting $dx_2=0$  in the first of the transformations to obtain 
 \be
 dt_1=\gamma_1\gamma_2(u_1u_2-1)dt_2.\label{eq:suppropercoord}
 \ee
 Similarly on setting $dt_1=0$ in the measurement to find $dt_2$  in terms of $dx_2$, the first equation gives the Lorentz-Fitzgerald contraction after some algebra as
 \be
 dx_1=\frac{dx_2}{\gamma_1\gamma_2(u_1u_2-1)}.\label{eq:supLF}
 \ee
  Relations from the perspective of $O_2$ are found by interchanging suscripts $1$ and $2$. If either observer approaches the speed of light from above, coordinate time diverges for finite proper time and coordinate length vanishes for finite proper length just as in the Lorentz transformation.  
   
  The  velocity transformation between these two observers follows  from the transformations as ($dx_i/dt_i$ is the velocity of an object relative to the $i^{th}$ supra-luminal observer)
  \be
  \frac{dx_1}{dt_1}=\frac{\frac{dx_2}{dt_2}+u_{rel}}{1+u_{rel}\frac{dx_2}{dt_2}},  ~~~~\frac{dx_2}{dt_2}=\frac{\frac{dx_1}{dt_1}-u_{rel}}{1-u_{rel}\frac{dx_1}{dt_1}}\label{eq:supveltrans}
  \ee
  which have the form of the  familiar velocity transformations, provided that the relative velocity is defined as 
 \be
 u_{rel}=\frac{u_2-u_1}{1-u_1u_2}.\label{eq:urel}
 \ee
 This `relative velocity'  is  both negative and reduced  in amplitude with respect to the normal relative velocity because  $u_1u_2>1$. The absolute  observer speeds  are always supra luminal in these expressions. 
 
 When $dx_2/dt_2=0$, we have $dx_1/dt_1=u_{rel}$, which goes to $1/u_2$ as $u_1\rightarrow \infty$ and $-1/u_1$ as $u_2\rightarrow\infty$. When $dx_1/dt_1=0$ we see that $dx_2/dt_2=-u_{rel}$. Thus  infinitely supra luminal observers perceive objects at rest relative to another supra luminal observer as being sub luminal, and to be at rest if the other observer also has infinite velocity. If one supra luminal observer perceives a photon, so also does the other whatever the relative speed. The speed of light continues to be an invariant.
 
 The expression that appears in the proper length and proper time transformations (equations (\ref{eq:suppropercoord}) and (\ref{eq:supLF}), namely 
 \be
 \gamma_1\gamma_2(u_1u_2-1)=\frac{1}{\sqrt{1-\frac{(u_1-u_2)^2}{(u_1u_2-1)^2}}}\equiv \gamma_{rel},\label{eq:gammarel}
 \ee
 as one would expect. 
 A peculiarity in $\gamma_{rel}$ is that if $u_2$ tends to infinity, $\gamma_{rel} \rightarrow \gamma_1 u_1$  and vice versa. This tends to unity if $u_1\rightarrow \infty$ also, as anticipated in the previous paragraph.
 
 We observe that, between supra luminal observers, the transformations  (\ref{eq:betu1u2}) are the normal Lorentz transformations in the form
 \bea
 t_1&=&\gamma_{rel}(t_2+u_{rel}x_2),\label{eq:betwu1u2t1}\\
 x_1&=&\gamma_{rel}(x_2+u_{rel}t_2),\label{eq:betwu1u2x1}
 \eea
 provided the relative velocity is given by equation (\ref{eq:urel}). The inverse transformations follow by interchanging $1$ and $2$, which also changes the sign of $u_{rel}$ as is normally the case.

 \section{Why should we care?}
 
 There are no reliable detections of supra luminal motion. However it is possible that we have not yet identified the  incriminating evidence. We proceed to examine various possibilities in sub sections.
 
  Our transformations require that there is a rotation of space-time  as the Cauchy horizon is crossed.  Any crossing  of the apparent horizon involves acting on an infinite mass/energy if a sub luminal  classical  particle with non-zero rest mass (often called  a `bradyon') were to achieve it. The same is true for a classical supra luminal massive particle ( a `tachyon')  crossing in the opposite sense. However there seems to be no objection in principle to the existence of a world `dual' to our own in which supra luminal matter has always existed \cite{BDS1962}. We have seen that the time and space axes of the two observers interchange \footnote{i.e. a full complex rotation  in complexified Minkowski space from space-like  to time-like  for each dimension}  as $u\rightarrow \infty$, which  does suggest a kind of `duality'. 
%We speculate further about  the permeability of the  horizon between the dual worlds based on quantum field ideas and `tunnelling'  subsequently in this discussion. 
 
 One may still ask whether  the existence of such a dual state has any relevance to the world we know. A well known peculiarity of tachyon signals is the ability to communicate with our past using the intermediary of another inertial observer (e.g. \cite{WR2006}), and the resulting consequences for causality. However such intra time  communication requires us to have experimental control of tachyons in both direction and speed. Such controls imply the ability to detect and manipulate tachyons, which is not yet possible. However as a general remark, the `block' picture of space time seems to  contain the past  in geometric form. Tachyons may be the necessary physical means to interact with it. The nature of such interaction must then  be restricted by a  consistency between the present and the past in a joint `reality'.  The geometry may not by equivalent to past events so that these can not be `rerun'.  
 
 \subsection{Can we detect them?}
 
   How might classical tachyons  be detected by a bradyon observer? Our arguments in this section continue to assume that the speed of light has a fundamental r\^ole in determining the metric of space-time, as in  determining the bradyon/tachyon boundary horizon.  If tachyons can either scatter or emit/absorb photons (i.e. if they are charged), the light signalling their presence  (in collision or otherwise) would be showing  a past state (more marked than the usual retarded state) of the tachyon particle.  Equation (\ref{eq:Inveltrans}) gives the apparent velocity of a tachyon {\it co-moving with $O'$ so that $dx'/dt'=0$}, equal to  $u>1$. 
  
  The most likely detection would be by a collision between a tachyon and a bradyon or a photon (if the tachyon is electrically  charged),  and so we  consider briefly the dynamics of a free tachyon, treating it initially as a classical particle. 
The `action' adopted by a sub luminal observer for  a free  supra luminal particle  of `asymptotic mass' (i.e. $u\rightarrow\infty$) $m>0$ might be expected as 
\be
S=m\int_1^2~\sqrt{dx^2-dt^2} \rightarrow m\int_1^2~dt\sqrt{ u^2-1}, \label{eq:action}
\ee
because the appropriate velocity is $u=dx/dt$, and the positive sign gives the {\it maximum action}  at $u=\infty$. The energy of a positive inertial mass tachyon calculated from this action is positive, $H=m/\sqrt{u^2-1}$, as is the momentum $mu/\sqrt{u^2-1}$ (which is $m$ in the asymptotic limit). With this choice of action the maximum occurs along the $x$ axis where $dt=0$, $H=0$ and $u\rightarrow\infty$. A free tachyon  does 'wish` to be at infinite velocity.

 Such particles with low energy ($u$ large)  would only be detectable in collisions with bradyons if the experiment were extremely sensitive to the energy and/or momentum balance. The momentum contribution can be considerable, being equal to $~mc$, depending on the tachyon `rest' mass. A finite momentum (depending on the tachyon mass) would  be missing in the collision balance, much as was the case with neutrinos. Rest masses of order the various neutrino masses would make the tachyons difficult to detect.
 
 Higher energy tachyons ($u\rightarrow 1+$) would be relatively obvious assuming a reasonable cross section for collision.
 However the supra luminal motion could make them seem unconnected to the site of the collision.  In any case there has to be a mechanism for accelerating tachyons that is more efficient than the gravitational field on known astronomical objects (see below). This may be difficult for neutral tachyons.

If the tachyon action were made to have a {\it minimum} at infinite velocity, then there would be a negative sign before the integral. This could be due to  negative rest mass. The minus sign without negative mass gives a negative energy\footnote{And a negative momentum}, which would have to be subtracted from the total energy of any interaction with a bradyon. The energy is still likely to be small.  The momentum would be antiparallel to the velocity and  would  create apparent missing momentum in a collision.  A negative sign with a negative rest mass returns us to the previous case except that $m$ would be the magnitude of the rest mass. %If the asymptotic mass were negative then the two negative signs would cancel. 

%Thus, either the  action (\ref{eq:action}) with positive mass and negative sign for minimization; or  simply a negative inertial mass, would show the anti-particle behaviour.  

 We  consider briefly  whether  the definition of inertial supra luminal reference frames can be different from those at  zero energy. Inertial observers are  freely falling observers in sub luminal General Relativity, but this will not suffice to define  very ($u\rightarrow\infty$) supra luminal inertial observers because in most cases they will not be in free fall (one only reaches light speed gravitationally at an event horizon). An alternative definition of an inertial observer is one in  uniform motion with respect to some standard of rest, which might be universal. The rest frame of the Higgs field comes to mind.  In the sub luminal regime, $u=0$ makes absolute sense only with respect to a standard inertial frame. Perhaps  in the end we must consider $u\rightarrow\infty$  as the supra luminal equivalent {\it hence only zero energy is absolute}. The limiting momentum $m$ gives   a reference by which to measure other momenta.
  
  All `preferred' supra luminal observers (i.e. inertial) would be uniformly moving with respect to any such  universal standard. Other supra luminal objects would have their instantaneous velocities transformed relative to  preferred observers $u_1$, or $u_2$ by equation (\ref{eq:supveltrans}), with say $dx_1/dt_1=v>1$ and $dx_2/dt_2=v'>1$.   A sub luminal observer will use  equation (\ref{eq:Inveltrans})   to obtain the  velocity $v$, which is seen by the preferred tachyon observer $u$ to be $v'$. If $uv'\ll 1$, we find that $v=v'+u$ and is therefore tachyonic. Should $uv'\gg 1$ however, we find $v=1/u+1/v'$, which is only tachyonic if $v'<1$. The inertial tachyon observer moves with velocity $u$ of course.
 
 Perhaps tachyons explain a current mystery without their identify being recognized. An obvious candidate is the dark matter and we turn to this in the next section.

 \subsection{Tachyons as Dark Matter}
 
  Astronomical dark matter has   properties that may correspond to tachyonic behaviour. It must be weakly interacting, which is achieved by low energy and low mass tachyons. It must be capable of gravitational lensing (which suggests positive mass energy), and it should encourage the growth of perturbations  in the early Universe. Low energy/momentum, `cold'  neutral tachyons  seem to fit these requirements. These objects would have substantial supra luminal velocity and small rest masses.  They would be essentially luminal particles if their total energy  $m_T$ (while still small relative to known masses) is much greater than their asymptotic mass/energy $m$. In fact $u^2=1+(m/m_T)^2$,  and in such a relatively high energy state they may be called `hot' for brevity.

For a perturbation of mass $m_p$ tachyons will be confined  gravitationally to a radius $R= G m_p \sqrt{u^2-1}/c^2$. This radius may be well outside the cosmological horizon and so we can expect $ m_p$ to grow beginning in the early Universe.

Tachyons could form galactic halos around a  spherical `seed' of mass $M(r)$ at radius $r$ by falling in from a distant low energy state.  Gravity will decelerate them until their energy attains $m_T=GM(r)m/rc^2$ and hence
 \be
  u(r)^2=1+(\frac{rc^2}{GM(r)})^2,\label{eq:tachyvel}
  \ee
  For a galactic size object with $M\approx 10^{11} M_\odot$ at $r_M\approx 10$  kpc one finds $u\approx 2\times 10^6 c$. Thus the energy remains small  and so does the momentum, $mc$, for small enough $m$ .  The tachyons will continue to be associated with the galaxy because the turn-round radius based on energy is  given by $R=GMu/c^2$, essentially $r_M$.  
  
  We note that  these numerical conclusions continue to hold at each radius $r$ in the galaxy if $M(r)\propto r$, so that $u\approx constant$. Only near an event horizon where $r\approx 2GM_\bullet/c^2$ 
  can $u$ aproach $\sqrt{5}$ under gravitational acceleration. The environments of collapsed objects may thus present anomalous supra luminal effects such as  tachyon outflow from the horizon. 
  
  %Thus, charged tachyons may produce anomalous  radiating motions near black holes, and uncharged tachyons can account for missing light. 
  
  Charged tachyons can produce Cherenkov radiation in a plasma and might be easily detectable.  Neutral tachyons in collision would be  difficult to detect, except as missing light. The only unique sign of a charged tachyon in a cloud or bubble chamber after an interaction that it survives, would be tracks of two identical particles moving away from the interaction.
  
  \subsection{Quantum Tachyons?}
  
% Our transformations define an apparent horizon passing through the origin of preferred sub luminal or supra luminal observers. 
 
 The apparent horizon is a direction in space time rather than a location and any parallel line is equivalent  If  tachyons are the excitations of quantum fields,  and if the horizon defines a natural vacuum, then pairs of tachyons and anti tachyons may be produced throughout any observer's space time, across the local horizon. 
 
 Such fields would contribute to the ground state of the vacuum. This state should have the invariant form $T_{ab}=-\rho_T\eta_{ab}$ for its stress/energy tensor, where $\eta$ is the usual Minkowski metric and $\rho_T$ is the tachyon energy density. We use a negative sign because we assume that $T_{ab}u^au^b>0$ for space-like $u^a$, but this is arbitrary. The energy density is zero for null $u^a$, which suggests considering the apparent horizon (given by the local null vector) as a natural ground state.

 If sub luminal particles are created with energy such that $u \le 1$ and supra luminal particles are created with $u\ge 1$, then it seems that quantum uncertainty may provide an effective flux of particles in both directions across the local apparent horizon. The minimum positional uncertainy for a relativistic particle is essentially the  de Broglie length $h/p$, and so from $\Delta p\Delta q\approx\hbar$ we see that the maximum $\Delta p\approx p$. This may permit tunnelling through the apparent horizon with nearly the same energy. This could not be classical tunnelling through a barrier because 
 the barrier height is actually infinite. Rather we may have to think of a bradyon and a tachyon becoming entangled  so that a velocity measurement may find either particle as either sub or supra luminal. 
 
 According to Feynman's interpretation (based on CPT invariance) of antiparticles as particles moving backwards in time, equation (\ref{eq:SL2})  indicates  that from the tachyon point of view, bradyons behave as antiparticles.  Similarly equation (\ref{eq:SLB2}) with $dx'=-udt'$ so that $dx=0$ indicates that tachyons are antiparticles for an inertial bradyon observer. This is another way of indicating partial duality between the two regimes. Complete duality would require every relativistic pair creation to produce a tachyon and a bradyon.
 
 However there are antiparticles produced in the sub luminal domain, independently of the existence of tachyons. That is, although tachyons may be antiparticles of  sub luminal particles and vice versa, not all antiparticles generated from the vacuum are tachyons.  Presumably, it is also the case  that not all antitachyons are bradyons.   
 
 There is nevertheless one interpretation that might restore the complete duality. Rather than identifying the vacuum with the apparent horizon, we might identify it with the sub luminal $u=0$ (implying a preferred frame of reference)  and the supra luminal $u\rightarrow \infty$. Then equation (\ref{eq:Inveltrans}) shows that we observe only sub luminal tachyon antiparticles as remarked at the end of section (\ref{sect:supraLorentz}). Similarly equation (\ref{eq:vtrans}) shows that near $u=0$ the tachyon observer sees only a bradyon antiparticle.
 
 Even with the aparent horizon as vacuum there is an approximate duality. 
 Equation (\ref{eq:Inveltrans})  with $dx'dt'=0$ shows that tachyon antiparticles produced near the $u=1$ vacuum state will not be very supra luminal for the bradyon observer. Similarly equation (\ref{eq:vtrans}) shows that 
 tachyon antiparticles will not be very sub luminal. Such discrepancies may not be  apparent unless measured carefully.
 
 %Relativistically, one of these may be the antiparticle of the other. In which case there would be an actual duality between the two domains.
 
 Crossing the apparent horizon (this becomes global as $x'\rightarrow t'$ according to equations (\ref{eq:SLB1}) and (\ref{eq:SLB2})) is similar to crossing an event horizon, except that it may not be unidirectional. An event horizon manifests the  problem of information loss (e.g. \cite{Hawk2014}, \cite{HPS2016}). Bradyon/tachyon pair productionnear the event horizon might allow the information to escape on the tachyon.
  
 If the true vacuum  of cosmic space-time is assumed to be comprised  predominantly of tachyons (assuming cancellation of other zero point fields perhaps due to super symmetry), then a crude estimate of the vacuum mass density is $\rho_T=3m_T/4\pi \lambda_D^3$.  
 Here we take  $\lambda_D$ to be the limiting De Broglie wavelength (also the Compton wavelength) $h/m_Tc$ for low energy tachyons. Then from 
 \be
 -\frac{\Lambda c^2}{8\pi G}\eta_{ab}=-\rho_T\eta_{ab},
 \ee
 we obtain $\Lambda=O(6m_T^4Gc/h^3)$.
 
  Setting this equal to the current value of $~1.11\times 10^{-56}$ $cm^{-2}$, requires $m=\approx 0.013$ eV as an upper limit to an average or dominant  tachyon rest mass.  Even if the tachyon energy is negative, we assume here that the rest mass is not.
This value is a possible neutrino mass, but neutrinos appear to be sub-luminal. 

A De Broglie wave length for a tachyon of this mass is $\approx 0.1mm$. In \cite{WR2006} it is noted that the  De Broglie wave phase velocity $w$ of a particle moving with speed $u$ satisfies $wu=c^2$. The De Broglie wave of  a  zero energy  tachyon particle would therefore be stationary in space as $u\rightarrow \infty$.  The wavelength  $\lambda_D= h/m_Tc$ in the limit.  As such a quantum tachyon, represented by its De Broglie wave eigenfunction, might be regarded as a property of the vacuum that has a  Compton  wavelength spatial periodicity.  This could be detectable using the Casimir effect.

\section{Conclusions/Discussion}

Our main result in this paper is a justification of the transformations (\ref{eq:SL1}), (\ref{eq:SL2}) and their inverses (\ref{eq:SLB1}) and (\ref{eq:SLB2}). 
These have been derived by extending a Lie motion across the local null surface that behaves like an apparent horizon. That is, space-like and time-like dimensions are interchanged on crossing this surface.  The transformation between sub luminal and supra luminal observers is not purely relative.

Recently a paper (\cite{DE2020}) has come to our attention. This paper makes an arbitrary choice of sign in their transformations to a supra luminal inertial observer that essentially identifies them with our transformations (\ref{eq:SL1}). It is unfortunate that a version of the preesent paper, submitted to the Royal Society in January of 2019, was not available to the authors of (\cite{DE2020}) because it would have added definiteness to their argument. 

I can not vouch for the original and  truly revolutionary ideas that they advocate concerning the origin of quantum mechanical concepts. One point does seem to me to be overlooked. That is, the time reversal viewed by a supra luminal observer. This seems to remove the multiple path argument (A  backwards in time to M and forwards in time to B) as seen in the supra luminal frame of reference. The argument for no strict causality is also affected negatively by time reversal.  In my argument the two orthogonal space coordinates remain unchanged,

The discussion in section (\ref{sect:suprasupratrans}) may be a little premature but it completes the story. It is useful if there are preferred supra luminal observers (e.g. zero energy) in combination with the velocity transformation (\ref{eq:vtrans}). 

In our extended discussion in the penultimate section, we have permitted several physical speculations in search of a method of detecting or recognizing tachyons. We have suggested that they may be relevant to both dark matter and dark energy. However these suggestions are only partly formulated and detailed work is required. If tachyons contribute to the vacuum energy, they may be detectable as De Broglie oscillations in the Casimir force as well as through the cosmological constant. In the latter connection a dominant mass of around $0.013$ eV is required.

The division of space-time into two domains by the apparent horizon cries out for some duality between the two regimes. We have suggested that this may occur near the horizon if one particle is the antiparticle of the other. Complete duality requires an absolute standard of rest in the sub luminal domain, corresponding to the zero energy infinite velocity standard, in the supra luminal domain.

  \section{acknowledgements}

%In the latter connection, a  tachyon energy of $\approx 1 GeV$ corresponds to a velocity of $\approx 10^{10}c$. The energy is essentially zero and the momentum is $m_T c$, small compared to the photon momentum. 

%Either condition (cold or hot) may comprise a low energy  distribution of dark matter that should react  to the initial perturbation spectrum. 

%This assumes that tachyons gravitate according to general relativity even at large $u$ where the energy is vanishing. Fortunately the momentum remains finite and will therefore contribute to the energy/momentum tensor. 

\label{lastpage}

\begin{thebibliography}{99}
\bibitem[\protect\citeauthoryear{Author}{2012}]{Author2012}
%starting Richard's
%\bibitem[\protect\citeauthoryear{Beck}{2016}]{Beck2016} Beck, R. 2016,\aap Review, {24},{\#4}
%\bibitem[\protect\citeauthoryear{Benjamin}{2012}]{B2012} Benjamin, R. A. 2012, {\it EDP Sciences}, EAS. {56}, {299}
%\bibitem[\protect\citeauthoryear{Carollo et al.}{2010}]{Car2010} Carollo, D., Beers, T., Chiba, M., Norris, J., Freeman, K., Lee, Y., Ivezi, Z., Rockosi, C. \& Yanny, B. 2010, \apj, {712}, {692}
%\bibitem[\protect\citeauthoryear{Alcubierre}{1994}]{Al1994} Alcubierre, M.,  1994,Class. Quantum Grav., {11}, {L73}
%\bibitem[\protect\citeauthoryear{Beck}{2015}]{Beck2015} Beck, Rainer, 2015, Astronomy \& Astrophysics Review, {24},{4}
%\bibitem[\protect\citeauthoryear{Blackman}{2015}]{Black2015} Blackman, E. G., 2015, Space Science Reviews, {188}, {59}
%\bibitem[\protect\citeauthoryear{Brandenburg et al.}{1992}]{BDMSST92} Brandenburg, A., Donnor, K. J., Moss, D., Shukurov, A., Sokoloff, D., Tuominen, I.,
%1992,\aap,{259},{453}
%\bibitem[\protect\citeauthoryear{Brandenburg}{2014}]{B2014} Brandenburg, A., 2014, {\it Simulations of Galactic Dynamos} in {\it Magnetic Fields in %Diffuse Media}, \# 407, {\it Astrophysics and Space Science Library}, 529
%\bibitem[\protect\citeauthoryear{Brentjens\& de Bruyn}{2005}]{BdeB2005}  Brentjens, M.A. \& de Bruyn, A.G., 2005, \aap, {441},{1217}
\bibitem[\protect\citeauthoryear{Bilaniuk, Deshpande \& Sudarshan}{1962}]{BDS1962} Bilaniuk, O.M., Deshpande, U.K. \& Sudarshan, E.C.G., 1962, Am. J. Phys., {30},{718}
\bibitem[\protect\citeauthoryear{Carter\&Henriksen}{1991}]{CH1991} Carter, B. \& Henriksen R. N., 1991, J. Math. Phys., {32(10)},{2580}
%\bibitem[\protect\citeauthoryear{Chyzy\&Buta}{2008}]{CB2008} Chyzy K.T.,\& Buta, R.J.,2008,\apj,{677},{L17}
%\bibitem[\protect\citeauthoryear{Damas-Segovia et al.}{2016}]{DamSegov2016} Damas-Segovia, A., et al., CHANG-ES VII, 2016, \apj, {824}, {30}
%\bibitem[\protect\citeauthoryear{de Vaucouleurs et al.}{1991}]{dev91} de Vaucouleurs, G., et al. 1991, Third Reference Catalogues of Bright Galaxies, Version 3.9, Springer-Verlag: New York
%\ bibitem[\protect\citeauthoryear{Durelle}{2011}]{Dur2011} Durelle, J. 2011, MSc. Thesis, Queen's University at Kingston, Ontario, Canada, ({\tt http://hdl.handle.net/1974/6940})
%\bibitem[\protect\citeauthoryear{Farrar}{2015}]{Gfarr2015} Farrar, G. 2015, {\it Astronomy in Focus}, vol. 1, XXIXth IAU General Assembly, Ed. Piero Benvenuti
%\bibitem[\protect\citeauthoryear{Ferri\`ere \& Terral}{2014}]{FT2014} Ferri\`ere, K., \& Terral, P. 2014, \aap, {561}, {100}
%\bibitem[\protect\citeauthoryear{Fraternali \& Binney}{2008}]{FB2008} Fraternali, F. \& Binney, J. 2008,\mnras,{386},{935}
%\bibitem[\protect\citeauthoryear{Greenspan}{1980}]{G1980} Greenspan, H. P. 1980, {\it The theory of rotating fluids}, paperback ed., Cambridge University Press, Cambridge, U.K.
%\bibitem[\protect\citeauthoryear{Gupta et al.}{2012}]{Gup2012} Gupta, A., Mathur, S., Krongold, K., Nicastro, F. \& Galeazzi, M. 2012, \apj, {756}, {L8}
%\bibitem[\protect\citeauthoryear{Hanasz et al.}{2009}]{Han2009} Hanasz, M., Woltanski, D., \& Kowalik, K. 2009, \apj, {706}, {L155} 
%\bibitem[\protect\citeauthoryear{Heald et al.}{2007}]{H2007} Heald, G., Rand, R., Benjamin, R., \& Bershady, M. 2007, \apj, {663}, {933}
%\bibitem[\protect\citeauthoryear{Henriksen\& Reinhardt}{1977}]{HR1977} Henriksen, R.N., \& Reinhardt, M. 1977, Astrophys. \& Sp. Sci, {49},{3}
%\bibitem[\protect\citeauthoryear{Henriksen}{2012}]{Hen2012} Henriksen, R. N. 2012, arxiv:1207.5430
%\bibitem[\protect\citeauthoryear{Gradshteyn\& Ryzhik}{1994}]{GR1994} Gradshteyn,I.S. \& Ryzhik, I.M.,1994, {\it Table of Integrals, Series, and Products}, Fifth Edition, Jeffrey, A. (Ed.), Academic Press, London
%\bibitem[\protect\citeauthoryear{Heald}{2009}]{Heald2009} Heald, G., 2009, {\it Cosmic Magnetic fields}, IAU Symposium, {259}, {591}, Strassmeier, K.G., Kosovichev,A.G. \& Beckman, J.E., Eds 
%\bibitem[\protect\citeauthoryear{Henriksen\&Reinhardt}{1977}]{HR1977} Henriksen, R.N. \& Reinhardt, M.,1977, Ap\&SS, {49},{3}
\bibitem[\protect\citeauthoryear{Dragan \& Ekert}{2020}]{DE2020} Dragan, A. \&Ekert, A.,2020, New J. Phys., {22},033038 
\bibitem[\protect\citeauthoryear{Feynman}{1949}]{Feyn1949} Feynman, R. P.,1949,Phys. Rev., {76}(6), {749}
%\bibitem[\protect\citeauthoryear{Foglizzo\& Henriksen}{1993}]{FH1993} Foglizzo,T. \& Henriksen, R.N., 1993, Phys. Rev. D, {48},{4645}
\bibitem[\protect\citeauthoryear{Hawking}{H2014}]{Hawk2014} Hawking, S.W., 2014, Arxiv:1401.5761v1{Hep-th}
\bibitem[\protect\citeauthoryear{Hawking, Perry \& Strominger}{2016}] {HPS2016} Hawking, S.W., Perry, M. \& Strominger A., 2016, Phys. Rev. Lett., {116},{231301}
\bibitem[\protect\citeauthoryear{Henriksen}{2011}]{Hen2011} Henriksen, R. N., 2011, {\it Practical Relativity},  John Wiley \& sons, Chichester, P019 8SQ, U.K.
\bibitem[\protect\citeauthoryear{Henriksen}{2015}]{Hen2015} Henriksen, R. N., 2015, {\it Scale Invariance: Self-Similarity of the Physical World}, Wiley-VCH, 69469 Weinheim,Germany
\bibitem[\protect\citeauthoryear{Hill and Cox}{2012}]{HC2012} Hill, J. M. \& Cox B.J., 2012, Proc. R. Soc. A, {468},{4174}
\bibitem[\protect\citeauthoryear{Hill and Cox}{2014}]{HC2014} Hill, J. M. \& Cox B.J., 2014, Z. Angew. Math. Phys., {65},{1251}
%\bibitem[\protect\citeauthoryear{Henriksen\&Irwin}{2016}]{HI2016} Henriksen, R.N. \& Irwin, J.A., 2015,\mnras,{458},{4210}
%\bibitem[\protect\citeauthoryear{Henriksen}{2017}]{Hen2017} Henriksen, R.N., 2017,http://arxiv.org/abs/1704.06954 
%\bibitem[\protect\citeauthoryear{Henriksen}{2017b}]{Hen2017b} Henriksen, R.N., 2017,\mnras,{469},{4806}
%\bibitem[\protect\citeauthoryear{Horellou\&Fletcher}{2014}]{HF2014} Horellou, Cathy \& Fletcher, A., 2014,\mnras,{441},{2049}
%\bibitem[\protect\citeauthoryear{Klein\& Fletcher}{2015}]{KF2015} Klein, U. \& Fletcher, A., 2015,{\it Galactic and Intergalactic Magnetic Fields}, Springer,
%Switzerland
%\bibitem[\protect\citeauthoryear{Kamphuis et al.}{2007}]{K2007} Kamphuis, P., Peletier, R., Dettmar, R.-J., van der Hulst, J., van der Kruit, P., \& Allen, R. 2007, \aap, {468}, {951}
%\bibitem[\protect\citeauthoryear{Kennedy}{2009}]{Ke2009} Kennedy, H. 2009, MSc Thesis, Queen's U. at Kingston, Ontario, Canada
%(%{\tt http://hdl.handle.net/1974/5134})
%\bibitem[\protect\citeauthoryear{Krause}{2009}]{Kr2009} Krause, M. 2009, Rev. Mex. AA, {36}, {25}
%\bibitem[\protect\citeauthoryear{Krause}{2015}]{Kr2015} Krause, M. 2015, {\it Highlights of Astronomy}, {16}, {399}
%\bibitem[\protect\citeauthoryear{Nagaosa\&Tokura}{2013}]{NT2013} Nagaosa, N. \& Tokura, Y., Nature Nanotechnology, {8},{899}
%\bibitem[\protect\citeauthoryear{Moffat}{1978}]{M1978} Moffat, H. K. 1978, {\it Magnetic field generation in electrically conducting fluids}, Cambridge University Press, Cambridge, U.K.
%\bibitem[\protect\citeauthoryear{Mora Partiarryo}{2016}]{CMP2016} Mora Partiarroyo, S. C., Ph.D. Thesis, Bonn Universit\"at, Bonn, Germany, http://hss.ulb.uni-bonn.de/2016/4537.htm
%\bibitem[\protect\citeauthoryear{Mora Partiarryo}{2017}]{CMP2017} Mora Partiarroyo, S. C., In preparation
%\bibitem[\protect\citeauthoryear{Moss\&Sokoloff}{2008}]{MS2008} Moss, D. \& Sokoloff,D. 2008, \aap, {487},{197}
%\bibitem[\protect\citeauthoryear{Moss et al.}{2015}]{M2015} Moss, D. Stepanov, R., Krause, M.,Beck, R., Sokolloff,D., 2015, \aap,{578}, {94}
%\bibitem[\protect\citeauthoryear{Landau \& Lifshitz}{1975}]{Lan1975} Landau, L., \& Lifshitz, E, 1975, {\it The Classical Theory of Fields},  Fourth Ed. Pergamon Press, Oxford, U.K.
%\bibitem[\protect\citeauthoryear{Li \& Wang}{2013}]{LiW2013} Li, J.-T., \& Wang, Q. D. 2013, \mnras, {428}, {2085}
%\bibitem[\protect\citeauthoryear{Marinacci et al.}{2010}]{M2010}Marinacci, F., Binney, J., Fraternali, F., Nipoti, C.,  Ciotti, L. \& Londrillo, P. 2010, \mnras, {404}, {1464}
%\bibitem[\protect\citeauthoryear{Marinacci et al.}{2011}]{M2011} Marinacci, F., Fraternali, F., Nipoti, C., Binney, J., Ciotti, L. \& Londrillo, P. 2011, \mnras, {415}, {1534}
%\bibitem[\protect\citeauthoryear{Oosterloo et al.}{2007}]{oos07}Oosterloo, T., Fraternali, F., \& Sancisi, R. 2007, \aj, 134, 1019
%\bibitem[\protect\citeauthoryear{Pakmor, Marinacci \& Springel}{2014}]{PMS2014} Pakmor, R., Marinacci, F., \& Springel, V. 2014, \apj, {783}, {L20}
%\bibitem[\protect\citeauthoryear{Rand}{2000}]{R2000} Rand, R. 2000, \apj, {537}, {L13}
\bibitem[\protect\citeauthoryear{Recami}{2001}]{Rec2001} Recami, E. 2001, Found. Phys., {31},{1119}
\bibitem[\protect\citeauthoryear{Rees}{1966}]{Rees66} Rees, M. J., Nature, {211},{468}
\bibitem[\protect\citeauthoryear{Rindler}{2006}]{WR2006} Rindler, W., 2006, {\it Relativity; Special, General and Cosmological}, 2nd Ed. Oxford University Press, Oxford

%\bibitem[\protect\citeauthoryear{Schmidt, Partiarroyo\& Krause}{2016}] {SPK2016}Schmidt, P.,Partiarroyo, S.C.M. \& Krause, M. 2016, Final Annual Meeting of the DFG research unit 1254, "Magnetization of Interstellar and Intergalactic Media",Berlin, October
%\bibitem[\protect\citeauthoryear{Sokoloff\&Shukurov}{1990}]{SS1990} Sokoloff, D. \& Shukurov, A., 1990, Nature, {347},{51}
%\bibitem[\protect\citeauthoryear{Sofue et al.}{1992}]{Sof1992} Sofue, Y., Reuter, H.-P., Krause, M., Wielebinski, R. \& Nakai, N. 1992, \apj, {395}, {126}
%\bibitem[\protect\citeauthoryear{T{\"u}llman et al.}{2000}]{T2000} T{\"u}llmann, R., Dettmar, R.-J., Soida, M., Urbanik, M. \& Rossa, J. 2000, \aap, {364},{36} 
%\bibitem[\protect\citeauthoryear{Sun}{2008}] {sun2008} Sun, X. H. 2008, \aap, {477}, {573}
\bibitem[\protect\citeauthoryear{Viera}{2012}]{V2012} Vierra, R.S., 2012, arxiv:1112.4187v2
%\bibitem[\protect\citeauthoryear{Wiegert et al.}{2015}]{WI2015}Wiegert, T., Irwin, J. A., Miskolczi, A., Schmidt, P., Mora, S. C., Damas-Segovia, A., Stein, Y., English, J., Rand, R. J., Santistevan, I., plus 14 co-authors 2015, \aj, {150}, {81}
%\bibitem[\protect\citeauthoryear{Zschaechner \& Rand}{2015a}]{ZRW15} Zschaechner, L., Rand, R., \& Walterbos, R. 2015a, \apj, {799}, {61}
%\bibitem[\protect\citeauthoryear{Zschaechner \& Rand}{2015b}]{ZR15} Zschaechner, L. \& Rand, R. 2015b, \apj, {808}, {153}
\end{thebibliography}
\end{document}